\documentclass[prl,twocolumn,aps,showpacs,superscriptaddress,reprint]{revtex4-1}
\usepackage{graphicx}
\usepackage{dcolumn}
\usepackage{bm}
\usepackage{epstopdf}
\usepackage{float}
\usepackage{amsmath,amsfonts}
\usepackage{xcolor}
\usepackage{comment}
\usepackage{hyperref}
\usepackage[normalem]{ulem}
\graphicspath{{./images/}}
\usepackage{pgfplots}

\usepackage{pgfplots}\pgfplotsset{compat=1.8}
\usepackage{tikz}
\usetikzlibrary{calc}
\usepackage{xcolor}

\begin{document}
\title{Ultra-strong trion-polaritons}

\author{M. A. Bastarrachea-Magnani}
\affiliation{Departamento de F\'isica, Universidad Aut\'onoma Metropolitana-Iztapalapa, Av. Ferrocarril San Rafael Atlixco 186, C.P. 09310 CDMX, Mexico.}
\email{bastarrachea@xanum.uam.mx}
\author{A. Camacho-Guardian}
\affiliation{Instituto de F\'isica, Universidad Nacional Aut\'onoma de M\'exico, Ciudad de M\'exico C.P. 04510, Mexico\looseness=-1}
\email{acamacho@fisica.unam.mx}

\begin{abstract} 
Due to the hybridization of charged excitons with cavity photons, trion-polaritons (TP) in microcavity semiconductors are a promising avenue for realizing strong polariton interactions and many-body polariton phases. We develop a quantum field theoretical formalism to study the formation of trion-polaritons in a microcavity semiconductor doped with itinerant electrons within the ultra-strong light-matter coupling regime. We predict a new class of many-body states resulting from the intriguing interplay of polarons and virtual photons in an uncharted territory of light-matter interactions and Feshbach physics.
\end{abstract}
\maketitle

{\it Introduction.-}
Polaritons are paradigmatic quasiparticles resulting from the strong coupling between matter excitations and light~\cite{Hopfield1958}. The strong coupling (SC) regime is achieved when the light-matter interaction exceeds the combined linewidth of the original constituents~\cite{Khitrova2006}. It is restricted to a light-matter coupling order much smaller than the bare energies of the excitations~\cite{Agarwal1971}. The SC regime has allowed for the observation of quantum coherent dynamics, which is the basis of current architectures in quantum information technologies, as well as intriguing and highly tunable light-matter states, such as slow light~\cite{Boller1991,Hau1999}, high- and room- temperature Bose-Einstein condensation~\cite{Amo2009,Deng2010,Kasprzak2006,Carusotto2013}, superfluidity~\cite{Sanvitto2011,Lerario2017}, topological photonics~\cite{Ozawa2019}, quantum vortices~\cite{Lagoudakis2008,Lagoudakis2009,Sanvitto2010}, polaritonic Feshbach resonances~\cite{Takemura2014,Navadeh2019}, and the promise of new platforms for strongly interacting polariton phases~\cite{Basov2021}. Particularly, itinerant electrons or holes in semiconductors can lead to the formation of a charged optical excitation, coined trion~\cite{Mak2013,Courtade2017,Emmanuele2020,Glazov2020}. Coupling the trion state to light can lead to the formation of trion-polaritons, which have received notable attention as they can be exploited to generate strongly correlated charged optical excitations~\cite{Sidler2017,Tan2020,efimkin2021electron,Li2021,Li2021a,Mulkerin2023,Tiene2023,Plyashechnik2023}, highly non-linear optical mediums~\cite{Koskal2021,Rana2021,Bastarrachea2021}, mechanisms to induce superconductivity~\cite{Julku2022,Milczewski2023arXiv,Zerba2023arXiv}, and to tune optical bi-stabilities and lasing~\cite{Wasak2021,Julku2021}. 

Increasing the light-matter coupling larger than the energy scales of the non-interacting system leads to the ultra-strong coupling (USC) regime. The USC has attracted attention in the last decades as it allows engineering the very nature of light and matter~\cite{FriskKockum2019,FornDiaz2019,MarquezPeraca2020}. In the USC regime, the ground state consists of an intrinsic squeezed vacuum combining matter and light features, reflecting the strongly correlated collective nature of the system~\cite{Ciuti2005,Ashhab2010,Felicetti2015}. The coupling of matter to vacuum fluctuations leads to counter-intuitive phenomena such as virtual photons dressing matter~\cite{SanchezMunoz2018} and the quest of extracting them~\cite{Ciuti2006,DeLiberato2007,DeLiberato2009,Auer2012,Garziano2013,Hagenmueller2016,Stassi2013,Huang2014,Lolli2015,Cirio2017}, observation of the dynamical Casimir effect~\cite{Wilson2011,Hagenmueller2016,Macri2018}, multiatom excitations via virtual photons~\cite{Garziano2016,Todorov2014,Huppert2016}, and the generation of photonic entangled states~\cite{Stassi2017,Macri2016,Macri2018,Garziano2015}. The strongly correlated state has sparked proposals on fast and protected quantum information processing protocols~\cite{Wang2010,Kyaw2015a,Romero2012,Wang2016,Kyam2015,Ofek2016,Schmidt-Kaler2003,Chow2012,Barends2014}, ultrafast qubit gates and memories~\cite{Romero2012,Nataf2011,Stassi2018}, quantum batteries~\cite{Ferraro2018,Ferraro2019,Ferraro2020}, non-linear optical processes~\cite{Kockum2017,Stassi2017}, and the modification of chemical reactions~\cite{Galego2015,Herrera2016}. In the field of microcavity semiconductors, the USC has nowadays promising avenues in intersubband polaritons~\cite{Anappara2007}, and organic polaritons~\cite{KenaCohen2013,Gambino2014,Gao2018}.

\begin{figure}[ht]
\includegraphics[width=0.97\columnwidth]{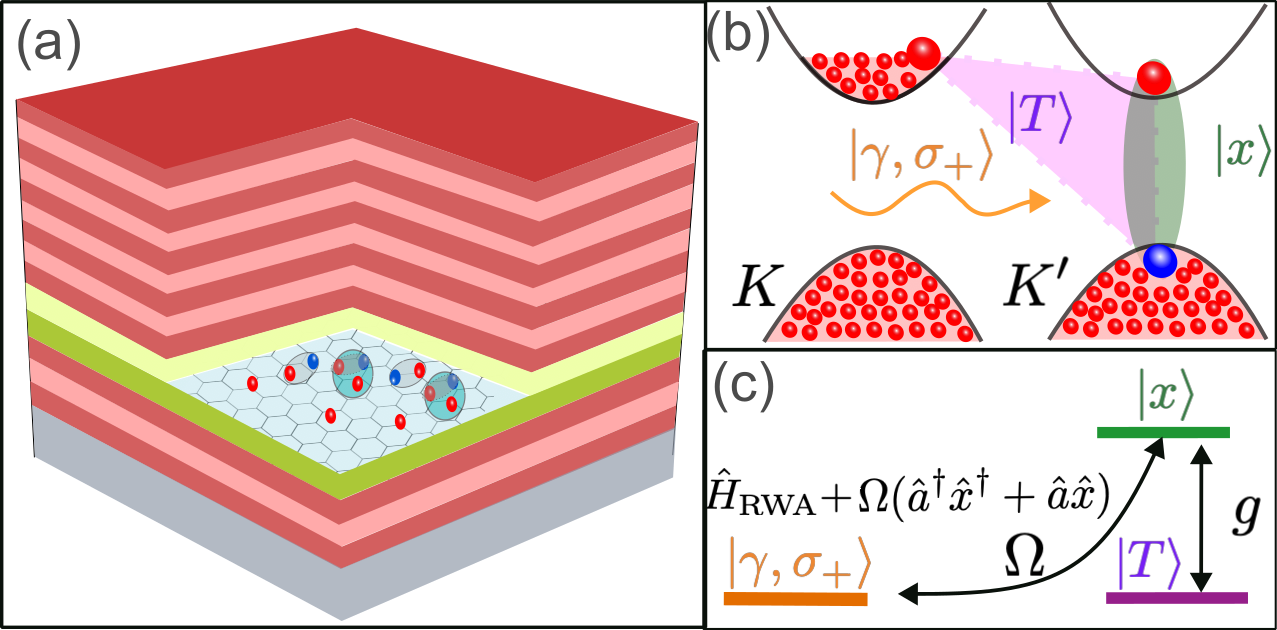}
\caption{(a) Schematic representation of the system: a 2D semiconductor doped with itinerant electrons and coupled to a high-finesse cavity. (b) Relevant transitions scheme: cavity photons coupled to excitons in the presence of a 2D Fermi gas. The exciton-electron interaction leads to a charged exciton-polariton, or {\it trion}. Light-matter coupling, including the counter-rotating processes from the USC regime, results in the formation of three exciton-polariton branches.}
    \label{fig:1}
\end{figure}

Inspired by the recent progress in the fields of microcavity semiconductors and strongly correlated polariton phases~\cite{Anantharaman2023}, in this Letter, we combine, at equal footing, the USC and Feshbach physics due to the strong interaction between microcavity exciton-polaritons and a two-dimensional electron gas (2DEG). Then, we predict the formation of charged optical excitations, coined trion-polaritons, in the USC regime and investigate the intriguing interplay between USC physics and many-body physics. To address the problem, we develop a diagrammatic theory that allows us to account for the Feshbach physics giving rise to the trion and the light-matter coupling in the USC regime. Understanding trion-polaritons may allow for new routes to the control of Feshbach resonances with electric fields~\cite{Schwartz2021,Wagner2023arXiv}, transport of optical excitations ~\cite{Cotlet2019,Chervy2020}, and new forms of optical bi-stabilities~\cite{Julku2021} beyond the SC regime. 

{\it System.-} 
Consider a two-dimensional semiconductor inside a high-finesse microcavity as schematically depicted in Fig.~\ref{fig:1} (a). Optical excitations in the semiconductor are described as excitons, which are strongly coupled to cavity photons 
\begin{gather}
\label{Hx}
\hat H_X=\sum_{\mathbf k}\left(\epsilon_{X\mathbf k}\hat x_{\mathbf k}^\dagger\hat x_{\mathbf k}+\omega_{\mathbf k}\hat a^\dagger_{\mathbf k}\hat a_{\mathbf k}\right)+ \\ \ \nonumber+
\sum_{\mathbf k}\Omega\left[\hat x_{\mathbf k}^\dagger \hat a_{\mathbf k}+\hat a_{\mathbf k}^\dagger \hat x_{\mathbf k}+ \xi\left(\hat x_{\mathbf k}^\dagger \hat a^\dagger_{-\mathbf k}+\hat a_{\mathbf k} \hat x_{-\mathbf k}\right)\right]+\\ \nonumber
\sum_{\mathbf k}D_{\mathbf k}\left(\hat a^\dagger_{\mathbf k}\hat a_{\mathbf k}+\hat a_{\mathbf k}\hat a^\dagger_{\mathbf k}+\hat a^\dagger_{\mathbf k}\hat a^\dagger_{-\mathbf k}+\hat a_{\mathbf k}\hat a_{-\mathbf k}\right)
\end{gather}
here $\hat x_{\mathbf k}^\dagger$ and $\hat a_{\mathbf k}^\dagger$ are the creation operators of an exciton and a cavity photon with in-plane momentum $\mathbf k$, respectively. The dispersion of the excitons and photons is given by $\epsilon_{X\mathbf k}=\omega_X+\mathbf{k}^2/2m_X$ and $\omega_{\mathbf k}=\omega_c+\mathbf{k}^2/2m_c$ where $m_{X/c}$ and $\omega_{X/c}$ are the respective masses and bare energies. The light-matter coupling is $\Omega$ with $\xi\in[0,1]$, a parameter tuning the strength of counter-rotating terms. We set the strength of the diamagnetic term to $D_{\mathbf k}\approx D=\Omega^2/\omega_X$, whose major effects is an energy shift due to an effective modification of the bare cavity photon's frequency~\cite{Ciuti2005,Garziano2020,Cortese2022}. One should notice that this term's presence and exact form depend on the particular system and electromagnetic gauge under study~\cite{MarquezPeraca2020}. In the SC $\Omega\ll\omega_{X}$, the rotating-wave approximation (RWA) allows for neglecting the counter-rotating terms due, so $\xi=0$. In our calculations, we take $\xi=1$ (Dicke limit).

In addition, itinerant electrons are driven into the system as a medium to create many-body polariton states, as in recent experiments~\cite{Sidler2017,Tan2020,Emmanuele2020}. Electrons interact with excitons via a short-ranged interaction and can form a bound state, the trion. The total Hamiltonian includes the electron's energy and electron-exciton interactions given by
\begin{gather}
\hat{H}_{e}=\sum_{\mathbf k}\epsilon_{f\mathbf k}\hat f^\dagger_{\mathbf k}\hat f_{\mathbf k}+\frac{g}{\mathcal A}\sum_{\mathbf q,\mathbf k,\mathbf k'}\hat x^\dagger_{\mathbf k+\mathbf q}\hat f^\dagger_{\mathbf k'-\mathbf q} \hat f_{\mathbf k'} \hat x_{\mathbf k} ,     
\end{gather}
where $\epsilon_{f\mathbf k}=\omega_{f}+k^2/2m_e$ denotes the electron dispersion, $m_e$ the electron mass and $\omega_{e}$ its energy at the bottom of the conduction band. The electrons form a two-dimensional electron gas (2DEG) with density $n_F,$ at zero temperature and a chemical potential set at the Fermi energy $\epsilon_F.$ $\mathcal{A}$ denotes the area of the system, and $g$ is the strength of the interaction between electrons and excitons, regarded as a contact interaction. This interaction supports a bound state  $\epsilon_{B\mathbf Q}=\omega_X+\omega_{f}-|\epsilon_B|+Q^2/2M$ where $\epsilon_B$ is the bare energy of the trion, $M=m_X+m_e,$ and $\mathbf Q$ is the total center-of-mass of the pair exciton-electron. The bound state leads to strong electron-polariton interactions due to polariton Feshbach physics ~\cite{Takemura2014,Navadeh2019,Schwartz2021}.

We employ a field theoretical approach based on the Green's function formalism to account for the combined strong matter-matter and ultra-strong light-matter interactions~\cite{Bruus2004}. For this purpose, we introduce a $4\text{x}4$ matrix Green's function defined as $\mathfrak{G}(\mathbf k,\tau)=-\langle T_{\tau}\{\psi_{\mathbf k}(\tau)\psi^\dagger_{\mathbf k}(0)\}\rangle,$ where $\psi_{\mathbf k}(\tau)=(\hat x_{\mathbf k},\hat x_{-\mathbf k}^{\dagger},\hat a_{\mathbf k},\hat a^\dagger_{-\mathbf k})^T$, and $T_\tau$ denotes the time-ordering operator. The Green's function obeys the Dyson's equation $\mathfrak{G}^{-1}(\mathbf k,\omega)=\mathfrak{G}^{-1}_0(\mathbf k,\omega)-\mathbf{\Sigma}(\mathbf k,\omega)$ where $\mathfrak{G}^{-1}_0(\mathbf k,\omega)$ is the ideal Green's function, $\mathbf{\Sigma}(\mathbf k,\omega)$ the self-energy which we discuss in detail in the following, and $\omega$ is the frequency.

{\it Theory of USC interacting polaritons.-} 
For clarity, we first study the Green's function for ideal polaritons in the USC regime in the absence of electrons. In this case, we have
\begin{gather} \nonumber
\mathfrak{G}_{0}^{-1}(\mathbf{k},\omega)=
\begin{bmatrix}
\mathcal{G}_{0X}^{-1}(\mathbf{k},\omega) & \mathbf{0}_{2}\\
\mathbf{0}_{2} & \mathcal{G}_{0c}^{-1}(\mathbf{k},\omega) 
\end{bmatrix},
\end{gather}
\begin{gather}\label{eq:s0}
\mathbf{\Sigma}_0(\mathbf k,\omega)=
\begin{bmatrix}
0& 0 & \Omega & \xi\Omega \\ 
0& 0 & \xi\Omega & \Omega \\ 
\Omega & \xi\Omega & 2D & 2D\\ 
\xi\Omega & \Omega & 2D & 2D \\ 
\end{bmatrix}.
\end{gather}
with $\mathcal{G}_{0i}^{-1}(\mathbf{k},\omega)$, $i=X,c$ the $2\text{x}2$ the bare exciton and photon $2\text{x}2$ Green's functions, and $\mathbf{\Sigma}_0(\mathbf{k},\omega)$  the $4\text{x}4$ self-energy that gives rise to the exciton-polaritons in the USC regime. In this case, $\Sigma_{13}=\Sigma_{31}=\Omega$ corresponds to the resonant terms,  whereas the anti-diagonal terms give the counter-rotating contributions. 

\begin{figure}[h!]
    \includegraphics[width=1\columnwidth]{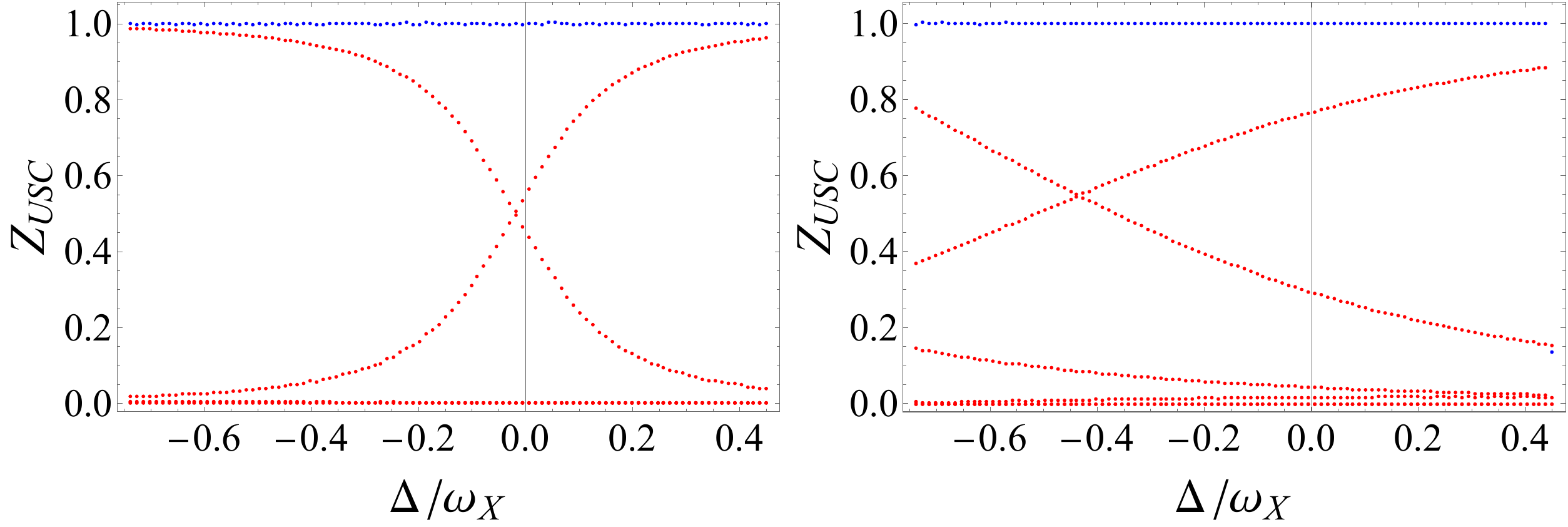}
    \caption{ The quasiparticle residue of the polariton branches at zero momentum for (a) $\Omega/\omega=0.1$ and (b) $\Omega/\omega=0.5$ (red dots). For clarity, we show $-Z_{3\mathbf{0}}$ and $-Z_{4\mathbf{0}}$. The normalization is always preserved $Z_{1\mathbf{0}}+Z_{2\mathbf{0}}-Z_{3\mathbf{0}}-Z_{4\mathbf{0}}=1$. Here we take $\xi=1$ and include the diamagnetic term.} 
  \label{FigZ}
\end{figure}

\begin{figure*}[ht]
\includegraphics[width=2\columnwidth]{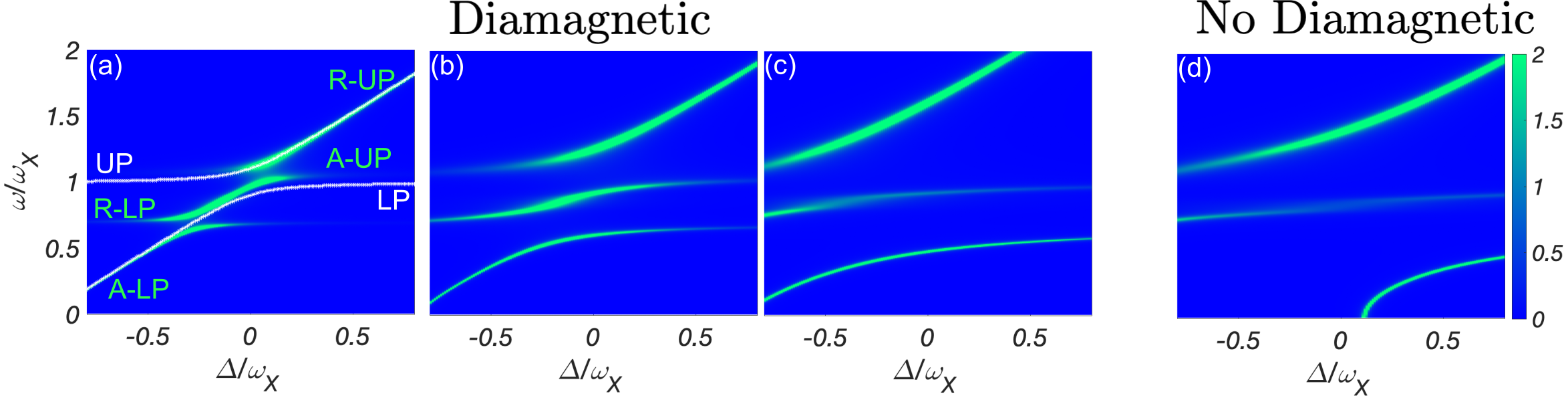}
\caption{Crossover from SC to USC trion-polaritons. Spectral function of cavity photons as a function of the light-matter detuning $\Delta$ and the frequency $\omega$ including the diamagnetic term $D$ for (a) $\Omega/\omega_X=0.1$, (b) $\Omega/\omega_X=0.25$, and (c) $\Omega/\omega_X=0.5$. Three polaron-polariton branches are observed. The lower (LP), middle (MP), and upper polariton (UP), which in the context of polaron-polariton physics, are coined as attractive lower-polariton (A-LP)(lower-polariton), repulsive upper polariton (R-LP) (upper polariton), while the middle polariton is formed from the repulsive lower-polariton (R-LP) branch which for positive detuning becomes the attractive upper polariton  (A-UP) state~\cite{Sidler2017,Bastarrachea2021,Bastarrachea2021b}. (d) Spectral function of cavity photons for a vanishing diamagnetic term $D=0$ for $\Omega/\omega_X=0.5$.}
    \label{Fig2}
\end{figure*}

An essential result of involving the counter-rotating process in the form of anomalous terms in Eq.~\ref{eq:s0} is the introduction of two new polariton branches that gain spectral weight as $\Omega/\omega_{X}$ increases. Within this approach, polariton branches emerge as poles of the Green's function. Within the RWA, the photonic Green's function $\mathfrak{G}_{33}(\mathbf k,\omega)$ has only two poles corresponding to the lower- and upper-polaritons. In the USC regime, the Green's function develops two additional poles. The quasiparticle residue $Z_{i\mathbf{k}}$ of the Green's function at the poles gives the spectral weight retained by the $i$-th polariton branch and equals the Hopfield coefficients. In Fig.~\ref{FigZ}, we show the quasiparticle residue of the polaritons as a function of the light-matter detuning. As expected, close to the SC regime, the Hopfield coefficients show only two polariton branches. As the light-matter coupling increases, the residues show two effects: a) The avoided crossing is shifted to negative detunings $\Delta$, mainly, due to the diamagnetic term and b) the two dominant branches yield spectral weight to the two emergent polariton states, which are a genuine USC effect. One should recall, that in the USC regime, the ground state contains a finite number of virtual photons, reflecting the strong-correlated light-matter state. This formalism, as we detail in the Supplemental Material~\cite{SM}, entirely agrees with the original formulation in Ref.~\cite{Hopfield1958,Quattropani1986} and the thorough studies in Refs.~\cite{Ciuti2005}.

The effects of the polariton-electrons interactions are included via an extended T-matrix approach~\cite{Schmidt2012,Bastarrachea2021} adapted to include the USC polaritons. This entails adding to $\mathbf{\Sigma}_0(\mathbf{k},\omega)$ a new self-energy term 
\begin{gather}
\mathbf{\Sigma}(\mathbf k,\omega)=
\begin{bmatrix}
\Sigma_{XX} &  0 & \Omega & \xi\Omega \\ 
0 & \Sigma_{XX} & \xi\Omega & \Omega \\ 
\Omega & \xi\Omega & 2D & 2D\\ 
\xi\Omega & \Omega & 2D & 2D \\ 
\end{bmatrix},
\end{gather}
with 
\begin{gather}
\Sigma_{XX}(\mathbf k,\omega)=\sum_{\mathbf q}n_{f}(\epsilon_{f\mathbf q})\mathcal{T}(\mathbf q+\mathbf k,\omega+\epsilon_{f\mathbf q}),
\end{gather}
where the T-matrix given by $\mathcal T^{-1}(\mathbf k,\omega)=g^{-1}-\Pi(\mathbf k,\omega)$ includes the many-body electron-exciton physics, and $n_{f}(x)=[1+\exp(\beta x)]^{-1}$ is the Fermi-Dirac distribution. 

In this case, the electron-exciton pair-propagator $\Pi(\mathbf k,\omega)$ is dressed by the cavity photons, as detailed in the Supp. Mat.~\cite{SM}. This approach considers the dressing of optical excitations by a 2DEG in the presence of a bound state (trion), thus it combines 2D Fermi-polaron physics~\cite{koschorreck2012attractive,Schmidt2012,Parish2013,levinsen2015strongly,klawunn2011fermi,Scazza2022} with USC polaritons~\cite{Ciuti2005}, and allows to study new physics beyond the RWA for SC polaron-polaritons or trion-polaritons~\cite{Sidler2017,pimenov2017fermi,Chang2019,Bastarrachea2021b}. 

To analyze the system, we focus on the spectral function of the cavity photons at zero in-plane momentum, defined as $A_{c}(\mathbf 0,\omega)=-2\text{Im}\mathfrak{G}_{33}(0,\omega)$. We consider a bare trion energy of $\epsilon_{B}=-25\text{meV}$~\cite{Mak2013}, effective exciton and cavity photon masses of $m_{X}=2m_{e}$ and $m_{c}=10^{-5}/m_{X}$. For concreteness, we take a fixed electron density ($n_{f}=1.3\text{x}10^{12}$) such that $\epsilon_{F}=0.25|\epsilon_{B}|$, where $\epsilon_{F}$ is the Fermi energy of the 2DEG, and study the system varying the ratio $\Omega/\omega_X$ and the cavity detuning from the exciton bare energy at zero in-plane momentum $\Delta=\omega_{c}-\omega_{X}$. 

{\it Polaron-polaritons in the USC regime.-}  The emergence of USC trion-polaritons is illustrated in Fig.~\ref{Fig2} as a function of the detuning $\Delta$, where we show the evolution of the spectral function as the ratio $\Omega/\omega_X$ is increased. In Fig.~\ref{Fig2} (a), for $\Omega/\omega_X=0.1,$ we observe that the spectral function closely resembles the experimental observation in Ref.~\cite{Sidler2017}. The signal arises from four distinct polaron-polariton branches coined attractive/repulsive lower/upper polaritons (labeled in the figure)~\cite{Bastarrachea2021b}. This regime can be well understood within the SC theory for polaritons as follows. In the SC polariton basis, the electron-polariton interaction gives a set of decoupled equations (when the scattering electron and a polariton do not change the polariton branch):
\begin{gather}
\label{EqSc}
\mathcal G^{-1}_{\text{LP}}(\mathbf 0,\omega)=\omega-\omega^{\text{SC}}_{\text{LP}}(\mathbf 0)-\mathcal C^2_{0}\Sigma_{XX}(\mathbf 0,\omega)\\ \nonumber
\mathcal G^{-1}_{\text{UP}}(\mathbf 0,\omega)=\omega-\omega^{\text{SC}}_{\text{UP}}(\mathbf 0)-\mathcal S^2_{0}\Sigma_{XX}(\mathbf 0,\omega),
\end{gather}
where $\omega^{\text{SC}}_{\text{LP}/\text{UP}}$ are the energy of the SC lower/upper polaritons, while the SC Hopfield coefficients at zero momentum are denoted by $\mathcal C^2_{0}$, $\mathcal S^2_{0}=1-\mathcal C^2_{0}$. From 2D polaron physics, it is well-established that the presence of a bound (trion) state gives rise to an attractive and repulsive polaron~\cite{Massignan2014}. Thus, each polaron branch individually couples to light, leading to two avoided crossings with a Rabi splitting given by $\Omega_{A/R}=Z_{A/R}\Omega$ for the attractive and repulsive polaron-polariton branches, where $Z_{A/R}$ are the attractive and repulsive polaron residues~\cite{Bastarrachea2021b}. One observes that for large negative/positive detuning, the polariton branches tend to be the attractive lower-polariton and the repulsive upper-polariton, respectively. The middle polariton branch acquires an S-shape as it is formed by the repulsive lower-polariton that transits to the attractive upper-polariton with increasing detuning. Increasing slightly the light-matter interactions ($\Omega/\omega_X=0.25$), the picture of the SC holds in Fig.~\ref{Fig2}(b). Larger light-matter coupling makes the two avoided crossings more evident. Furthermore, the middle polariton retains its characteristic S-shape, which is shifted to negative detuning.  

Driving further the system into the USC regime leads to a dramatic change in the quasiparticle properties of the polaritons, as shown in Fig.~\ref{Fig2} (c) for $\Omega/\omega_{X}=0.5$. The S-shape of the middle polariton is barely visible, as the two avoided crossings are now shifted and cannot be completely distinguished. Notably, as the system enters the USC regime, the S-shape of the middle polariton shrinks because the polaron-polariton branches are being dressed by the virtual photons from the two new emerging polariton branches. 

The competition between the light-matter and exciton-electron interactions helps us understand the middle branch's robustness. By increasing $\Omega/\omega_{X}$ for fixed $|\epsilon_{B}|$ in Figs.~\ref{Fig2}(a)-(c) we are effectively decreasing the ratio $|\epsilon_B|/\Omega=2.5, 1,$ and $0.5$. Then, the middle polariton branch is more robust against large light-matter couplings when the Feshbach physics and the Rabi coupling are comparable  $|\epsilon_B|/\Omega\sim 1$.

\begin{figure}[h!]
    \includegraphics[width=1\columnwidth]{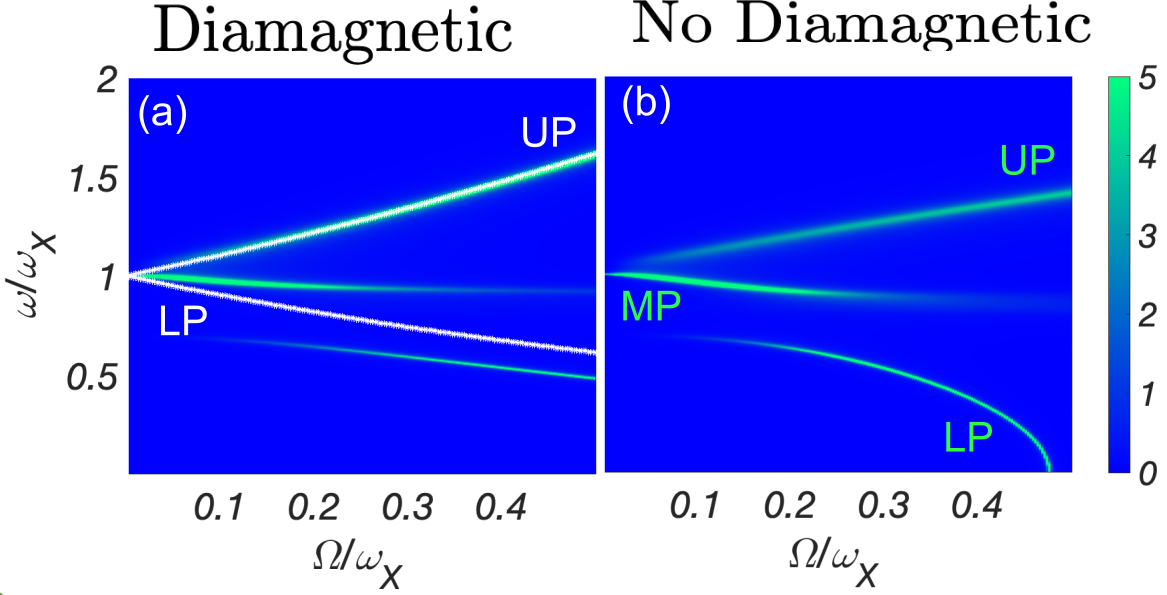}
    \caption{Spectral function of the cavity photons as a function of $\Omega/\omega_X$ and $\omega$. The white dots give the energy of the exciton-polaritons in the absence of the 2DEG.} 
  \label{Fig3}
\end{figure}

To unravel the distinct roles played by the counter-rotating terms and the diamagnetic contribution, we show in Fig.~\ref{Fig2}(b) the spectral function of the cavity photons setting $D=0$. We observe that the diamagnetic term only has qualitative effects on the middle and upper polariton state but dramatically impacts the lower polariton branch. Without the diamagnetic term, the lower-polariton branch goes to negative energies, signaling the onset of virtual photons in the system. Instead, the diamagnetic term shifts the branch back to positive energies, which entails an energy cost for the macroscopic population of virtual excitations~\cite{Larson2021}. Hence, both the counter-rotating and diamagnetic terms play a crucial role in the formation of USC trion-polaritons.

The intriguing interplay between polaron and polariton physics occurring in the crossover from SC to USC trion-polaritons is further unveiled in Fig.~\ref{Fig3}(a) where we show the spectral function as a function of $\Omega/\omega_X$ for $\delta=0$. In stark contrast to polaritons in the absence of the 2DEG (white dots), not all polariton branches depart from the bare exciton energy for $\Omega/\omega_X\ll 1$. The lower polaron-polariton energy is visibly shifted to lower energies, and with increasing the Rabi coupling, it remains significantly distanced from the polaritons in the absence of 2DEG. Furthermore, while this branch starts with a vanishing spectral weight for small $\Omega,$ as the light-matter interaction increases, it gains it. This reflects both real and virtual photons dressing the attractive polaron-polariton. The lower polariton parametric evolution as a function of the detuning is smoother, suggesting not only an expected stronger correlation but also a region where the light-matter hybridization is more robust. 

On the other hand, the middle polariton takes most of the spectral weight close to the SC regime ($\Omega/\omega_{X}$). This branch cedes spectral weight when driving the system to the USC but remains distinguishable. We stress that the middle polariton branch is a unique feature that results from the strong coupling to the trion state, separating the light-matter coupling between the attractive and repulsive polarons. Finally, the upper polaron-polariton emerges blurred deep in the SC regime and gains spectral weight with increasing $\Omega$. It represents a {\it weakly} light-matter interacting state.  One could naively assume that the fading of the middle polaron-polariton branch implies the effects of strong exciton-electron interactions become negligible in this regime. Instead, we observe the lower polaron-polariton gains significant spectral weight. Since in the USC regime, the lower polaron-polariton branch is visibly separated from the USC polaritons in the absence of 2DEG, it retains trion-physics, as previously discussed, we interpret this as a persistent imprint of Feshbach physics in the regime where virtual photon processes become relevant. Finally, in Fig.~\ref{Fig3}(b), we show the evolution of the trion-polaritons as a function of $\Omega$ with vanishing diamagnetic term. Here again, we observe an interplay between the counter-rotating and diamagnetic terms, the latter allowing the lower polariton to retain its quasiparticle character. 

{\it Discussion and Outlook.-} 
Trion-polaritons have received notable attention given breakthrough experiments which have unfolded novel physics and potential applications at the single particle level~\cite{Sidler2017} and to engineer highly non-linear optical mediums ~\cite{Tan2020,Emmanuele2020,efimkin2017many,efimkin2018exciton,efimkin2021electron,rana2020many,Koskal2021,Rana2021,Bastarrachea2021,Li2021,Li2021a,Mulkerin2023,Tiene2023,Plyashechnik2023}. Due to the inherent nature of the state-of-art experiments, the theoretical studies have been devoted to the characterization of trion-polaritons in the SC regime, leaving the interplay between USC and Feshbach physics completely unexplored.
  
In this Letter, we have studied USC polaritons strongly coupled to free carriers and shown the emergence of USC trion-polaritons, which can also be understood in terms of USC polaron-polaritons. For this purpose, we developed a diagrammatic approach that allowed us to systematically combine the description of the USC polaritons beyond the RWA and the underlying Feshbach physics that gives rise to the trion. We demonstrated that the imprints of the trion state on the optical response of the system persist in the USC regime and that the features of the trion-polaritons are heavily modified in the USC regime compared to SC trion-polaritons, where we conjecture the new polaron-polariton states are dressed by virtual photons. With the recent progress in increasing light-matter coupling strength in transition metal dichalcogenides (TMD)~\cite{Anantharaman2023}, the possibility of observing USC polaron-polariton states via doping itinerant electron~\cite{Tan2020,Emmanuele2020} seems at hand. 
Our theoretical framework opens up an avenue to further studies on strongly interacting polaritons in the USC regime, which may unfold new opportunities to design multiplatform polariton-polariton interactions~\cite{Bastarrachea2021}, few-body states of polaritons~\cite{Levinsen2019,bastarrachea2019strong,Camacho2021,camacho2022strong,kudlis2024theory}, optical control with electric fields~\cite{Cotlet2019,Chervy2020} beyond the well-established SC limit, and to study of virtual photons. 

{\it Acknowledgments.-} A. C. G. acknowledges financial support from Grant UNAM DGAPA PAPIIT No. IA101923, PAPIME No. PE101223 and PIIF 2023. 
\bibliography{references}

\setcounter{section}{0}
\setcounter{equation}{0}
\setcounter{figure}{0}
\setcounter{table}{0}
\setcounter{page}{1}
\makeatletter
\renewcommand{\theequation}{S\arabic{equation}}
\renewcommand{\thefigure}{S\arabic{figure}}
\renewcommand{\bibnumfmt}[1]{[S#1]}
\renewcommand{\citenumfont}[1]{S#1}

\clearpage

\onecolumngrid
\begin{widetext}

\begin{center}
\section{Supplemental Material: Ultra-strong trion-polaritons}
\end{center}

\subsection{Non-Interacting USC polaritons}

We start detailing the field theoretical approach to treat the formation of polaritons in the ultra-strong coupling (USC) without strong matter interactions. Our many-body formalism is based on the imaginary-time Green's function $\mathfrak{G}(\mathbf{k},\tau)$, where $\mathbf{k}$ is the in-plane momentum and $\tau$ the imaginary-time~\cite{Fetter1971,Bruus2004}. In this case, we have
\begin{gather}
\mathfrak{G}(\mathbf k,\tau)=-\langle T_{\tau}\{\hat{\psi}_{\mathbf k}(\tau)\hat{\psi}^\dagger_{\mathbf k}(0)\}\rangle,
\end{gather}
where the $\hat{\psi}_{\mathbf k}(\tau)$ is the field operator including the four relevant fields, i.e., $\psi_{\mathbf k}(\tau)=(\hat x_{\mathbf k},\hat x_{-\mathbf k}^{\dagger},\hat a_{\mathbf k},\hat a^\dagger_{-\mathbf k})^\text{T},$ with $T_\tau$ denotes the time-ordering operator, and $\{\cdot,\cdot\}$ the anticommutator. Here, $\hat a_{\mathbf k}$ and $\hat x_{\mathbf k}$ are the annihilation operators of the cavity photon and exciton fields, respectively. The Green's function is a $4\text{x}4$ matrix that, in the absence of electron doping, can be obtained exactly as $\mathfrak{G}^{-1}(\mathbf{k},\omega)=\mathfrak{G}_{0}^{-1}(\mathbf{k},\omega)-\mathbf{\Sigma}(\mathbf{k},\omega)$, where $\mathfrak{G}_{0}^{-1}(\mathbf{k},\omega)$ is the free Green's function given by
\begin{gather} \nonumber
\mathfrak{G}_{0}^{-1}(\mathbf{k},\omega)=
\begin{bmatrix}
\mathcal{G}_{0X}^{-1}(\mathbf{k},\omega) & \mathbf{0}_{2}\\
\mathbf{0}_{2} & \mathcal{G}_{0c}^{-1}(\mathbf{k},\omega) 
\end{bmatrix}=
\begin{bmatrix}
\omega-\epsilon_{X\mathbf k} & 0 & 0 & 0 \\ 
0&-\omega-\epsilon_{X\mathbf k} & 0& 0 \\ 
0 & 0 & \omega-\omega_{\mathbf k} & 0\\ 
0 & 0 &0 & -\omega-\omega_{\mathbf k} 
\end{bmatrix}. 
\end{gather}
Here, $\mathcal{G}_{0i}^{-1}(\mathbf{k},\omega)$ with $i=X,c$ is $2\text{x}2$ matrix representing the bare exciton and photon propagators, that is, for vanishing light-matter coupling. Interactions are encoded in the $4\text{x}4$ self-energy matrix, given by
\begin{gather}
\mathbf{\Sigma}_{0}(\mathbf{k},\omega)=
\begin{bmatrix}
\mathbf{0}_{2} & \Sigma_{\Omega}(\mathbf{k},\omega)\\
\Sigma^{*}_{\Omega}(\mathbf{k},\omega) & \Sigma_{D}(\mathbf{k},\omega)
\end{bmatrix}=
\begin{bmatrix}
0& 0 & \Omega & \xi\Omega \\ 
0& 0 & \xi\Omega & \Omega\\ 
\Omega & \xi\Omega & 2D_{\mathbf{k}} & 2D_{\mathbf{k}}\\ 
\xi\Omega & \Omega & 2D_{\mathbf{k}} & 2D_{\mathbf{k}} \\ 
\end{bmatrix}.
\end{gather}
where $\Sigma_{\Omega}(\mathbf{k},\omega)=\Sigma^{*}_{\Omega}(\mathbf{k},\omega)=\Omega\left(\mathbb{I}_{2}+\xi\mathbb{J}_{2}\right)$ encodes the light-matter interactions beyond the rotating-wave approximation (RWA), with $\mathbb{I}_{2}$ and $\mathbb{J}_{2}$ the $2\text{x}2$ identity and exchange matrices, and $\Sigma_{D}(\mathbf{k},\omega)=2D_{\mathbf{k}}\mathbb{I}_{2}$ the diamagnetic coupling. As described in the main text, we set $D_{\mathbf{k}}=D=\Omega^{2}/\omega_{x}$~\cite{Ciuti2005,Garziano2020}. The relevant self-energy terms can be pictured in their diagrammatic representation as in Fig.~\ref{fig:s1}.

\begin{figure*}[ht]
\includegraphics[width=0.7\columnwidth]{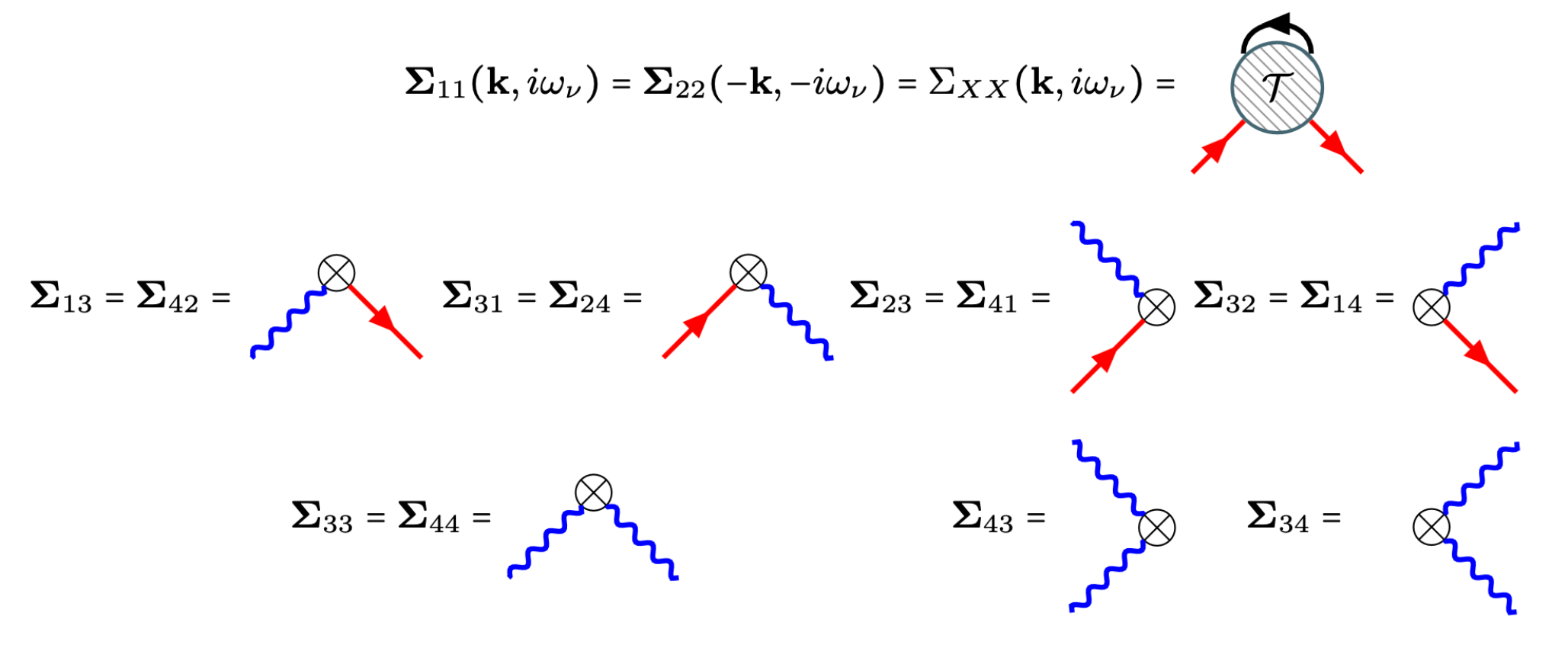}
\caption{Diagrammatic representation of the relevant self-energy matrix elements. The crossed dot denotes the Rabi splitting $\Omega$, the solid red line is the exciton propagator, the blue wavy line is the photon propagator, whereas the solid black line gives the electron propagator. The big circle corresponds to the exciton-electron T-matrix.}
    \label{fig:s1}
\end{figure*}

In the SC regime, we can discard the counter-rotating terms of the light-matter interaction $(\xi=0)$ and the diamagnetic terms via the RWA, so $\Sigma_{\Omega}(\mathbf{k},\omega)=\Omega\mathbb{I}_{2}$, $\Sigma_{D}(\mathbf{k},\omega)=0$, and 
\begin{gather}
\mathbf{\Sigma}_{\text{RWA}}(\mathbf{k},\omega)=\Omega
\begin{bmatrix}
\mathbf{0}_{2} & \mathbb{I}_{2}\\
\mathbb{I}_{2} & \mathbf{0}_{2}
\end{bmatrix}
\end{gather}
In this regime, one obtains the standard SC polaritons. Their quasiparticle properties (energy and residue) are obtained from the photon and exciton Greens' functions. Since the Green's function is decomposed into two independent $2\text{x}2$ blocks, we get that for the exciton Green's function
\begin{gather}
\mathcal \mathcal{G}^{-1}_{XX}(\mathbf k,\omega)=\mathfrak{G}^{-1}_{0XX}(\mathbf k,\omega)=\omega-\epsilon_{X\mathbf k}-\Omega^2\left(\omega-\omega_{\mathbf k}\right)^{-1},    
\end{gather}
which only has two poles, at the energies of the SC lower and upper polaritons
 \begin{gather}
 \label{SCener}
\omega_{\text{LP/UP}\mathbf{k}}^{\text{SC}}=\frac{1}{2}\left(2\epsilon_{X\mathbf k} +\Delta_{\mathbf k} \pm\sqrt{\Delta_{\mathbf{k}}^2+4\Omega^2 } \right),
\end{gather}
where $\Delta_{\mathbf{k}}=\epsilon_{X\mathbf k} -\omega_{\mathbf k}$ is the light-matter detuning at momentum $\mathbf{k}$. In particular, at zero detuning (maximum hybridization) and for $\mathbf k=0$, we have $\omega_{\text{LP,UP}}^{\text{SC}}=\omega_X\pm\Omega$. In addition, the quasiparticle residue of each polariton relates directly to the ideal Hopfield coefficients
\begin{gather}
Z_{\text{LP}\mathbf{k}}^{\text{SC}}=\mathcal C_{\mathbf{k}}^2=\frac{1}{2}\left(1+\frac{\epsilon_{X\mathbf k} -\omega_{\mathbf k}}{\sqrt{(\epsilon_{X\mathbf k} -\omega_{\mathbf k})^2+4\Omega^2 } }\right),\;\;\;
Z_{\text{UP}\mathbf{k}}^{\text{SC}}=\mathcal S_{\mathbf{k}}^2=\frac{1}{2}\left(1-\frac{\epsilon_{X\mathbf k} -\omega_{\mathbf k}}{\sqrt{(\epsilon_{X\mathbf k} -\omega_{\mathbf k})^2+4\Omega^2 } }\right), 
\end{gather}
with $Z_{\text{UP}\mathbf{k}}^{\text{SC}}+Z_{\text{UP}\mathbf{k}}^{\text{SC}}=\mathcal C_{\mathbf{k}}^2+\mathcal S_{\mathbf{k}}^2=1$. 

Beyond the RWA, the $2\text{x}2$ matrices couple via the counter-rotating terms, and the Green's function, even in the absence of electron doping, becomes more complex. A simple check for zero-momentum and vanishing detuning gives for the polariton energies
\begin{gather}
\omega_{\text{LP/UP}\mathbf{k}=0}^{\text{USC}}=\pm\left(\Omega\mp \omega_{X}\sqrt{1+\frac{\Omega^{2}}{\omega_{X}^{2}}}\right),
\end{gather}
in agreement with Refs.~\cite{Ciuti2005,Garziano2020}. In the $\Omega/\omega_X\ll 1$ limit, we recover the energies of the lower and upper-polaritons in Eq.~\ref{SCener}.  In contrast to the Green's function under the RWA, in the USC regime, it acquires four poles. The energy of the quasiparticle branches for arbitrary in-plane momentum and detuning is more complex. The USC polariton dispersion relationships read as
\begin{gather}
\omega_{\text{LP}\mathbf{k}}^{\text{USC}}=\frac{1}{\sqrt{2}}\left\{\frac{}{}\epsilon_{X\mathbf{k}}^2+\omega_{\mathbf{k}}^{2}+4D_{\mathbf{k}}\omega_{\mathbf{k}}+2 \left(1-\xi\right)\left(1+\xi\right) \Omega ^2\right.
\nonumber\\ \nonumber
-\left.\sqrt{\left(
\omega_{\mathbf{k}}^{2}+4D_{\mathbf{k}}\omega_{\mathbf{k}}-\epsilon_{X\mathbf{k}}^2
\right)^{2}-4 \Omega^{2}
\left[(1+\xi)\omega_{\mathbf{k}}+(1-\xi)\epsilon_{X\mathbf{k}}\right]
\left[(1-\xi )\left(\omega_{\mathbf{k}}+4D_{\mathbf{k}}\right)-(1+\xi )\epsilon_{X\mathbf{k}}\right]
}\right\}^{\frac{1}{2}}.
\\
\omega_{\text{UP}\mathbf{k}}^{\text{USC}}=\frac{1}{\sqrt{2}}\left\{\frac{}{}\epsilon_{X\mathbf{k}}^2+\omega_{\mathbf{k}}^{2}+4D_{\mathbf{k}}\omega_{\mathbf{k}}+2 \left(1-\xi\right)\left(1+\xi\right) \Omega ^2\right.
\nonumber\\ \nonumber
+\left.\sqrt{\left(
\omega_{\mathbf{k}}^{2}+4D_{\mathbf{k}}\omega_{\mathbf{k}}-\epsilon_{X\mathbf{k}}^2
\right)^{2}-4 \Omega^{2}
\left[(1+\xi)\omega_{\mathbf{k}}+(1-\xi)\epsilon_{X\mathbf{k}}\right]
\left[(1-\xi )\left(\omega_{\mathbf{k}}+4D_{\mathbf{k}}\right)-(1+\xi )\epsilon_{X\mathbf{k}}\right]
}\right\}^{\frac{1}{2}}.
\end{gather}
The exciton's Green's function takes the form of
\begin{gather}
\label{fullG2}
G_{X}(\mathbf{k},\omega)=\frac{Z_{1}^{\text{USC}}(\mathbf{k})}{\omega-\omega_{\text{LP}}^{\text{USC}}(\mathbf k)}+\frac{Z_{2}^{\text{USC}}(\mathbf{k})}{\omega-\omega_{\text{UP}}^{\text{USC}}(\mathbf{k})}
-\frac{Z_{3}^{\text{USC}}(\mathbf{k})}{\omega+\omega_{\text{LP}}^{\text{USC}}(\mathbf{k})}-\frac{Z_{4}^{\text{USC}}(\mathbf{k})}{\omega+\omega_{\text{UP}}^{\text{USC}}(\mathbf{k})}, 
\end{gather}
with $Z_{1}^{\text{USC}}+Z_{2}^{\text{USC}}-Z_{3}^{\text{USC}}-Z_{4}^{\text{USC}}=1$, the normalization condition over the USC Hopfield coefficients $Z_{i\mathbf{k}}^{\text{USC}}$. Again, for small $\Omega/\omega_X$, we expect two dominant poles with energies tending to those of the SC polariton energies, with their quasiparticle residues going to the values of the ideal SC Hopfield coefficients. The other two poles will have negligible residues. 

\subsection{USC Polariton-Electron Scattering}

Here, we offer details on the polariton-electron scattering beyond the RWA to understand how the polaron effects change from the SC to the USC regimes. For this purpose, we analyze the many-body scattering matrix in the ladder approximation~\cite{Fetter1971,Bruus2004,Bastarrachea2021}. As it is described in the main text, it reads
\begin{gather}
\mathcal{T}^{-1}(\mathbf{k},i\omega_{\nu})=g^{-1}-\Pi(\mathbf k,i\omega_{\nu}),
\end{gather}
where $i\omega_{\nu}$ is a fermionic Matsubara frequency with $\nu=2\pi\nu/\beta$, $\nu=0\pm1,\pm2,...$, $\beta$ is the inverse of temperature, $g$ is the two-body electron-exciton contact interaction, and the electron-exciton pair-propagator is given by
\begin{gather}
\Pi(\mathbf{k},i\omega_{\nu})=-\frac{1}{\beta\mathcal{A}}\sum_{\mathbf q,i\kappa_\nu}G_{0f}(-\mathbf q,-i\kappa_\nu)G_{X}(\mathbf k+\mathbf q,\kappa_{\nu}+i\omega_{\nu}),   
\end{gather}
where $G_{0f}^{-1}(\mathbf{k},i\kappa_{\nu})=i\kappa_{\nu}-\epsilon_{f\mathbf{k}}-\epsilon_{F}$ is the bare electron propagator, $\epsilon_{F}$ is the Fermi energy of the two-dimensional electron gas, and the sum is made over both momenta $\mathbf{k}$ and fermionic matsubara frequencies $i\kappa_{\nu}$. $\mathcal{A}$ is the area of the system. The exciton self-energy is calculated as in Eq.~(4) from the main text, and the final Green's function is obtained by analytic continuation $\mathfrak{G}(\mathbf{k},\omega)=\left.\mathfrak{G}(\mathbf{k},i\omega_{\nu})\right|_{i\omega_{\nu}\rightarrow\omega+i0^{+}}$. To account for the light-matter interaction at the USC level, we employ the dressed exciton propagator as $\mathcal{G}_{X}(\mathbf{k},\omega)=\mathfrak{G}_{XX}(\mathbf k,\omega)$ with  $\mathcal{G}_{XX}(\mathbf{k},\omega)$ calculated from Eq.~\ref{fullG2}. We assume that the presence of the polaritons does not perturb the 2DEG, so we evaluate the scattering matrix in the limit of vanishing electron density but retaining the full light-matter coupling. 

Figure~\ref{fig:s3} shows the real and imaginary parts of the scattering matrix $\mathcal{T}$ for $\Omega/\omega_X=0.5$ and $\Delta=0$ for $\epsilon_B/\epsilon_X=-0.1$.  Solutions using the full exciton Green's function in Eq.~\ref{fullG2} are illustrated by the blue and red dots for the real and imaginary parts of the scattering matrix, respectively. For comparison, we show the scattering matrix without light-matter coupling $\Omega/\omega_X=0$ (solid lines), following the same color coding as before. We observe a complete agreement with and without light-matter interactions, which indeed lie on top of each other. Physically, this can be understood as follows: At zero detuning, cavity photons couple to excitons as long as the in-plane momentum $k_{\Omega}\leq \sqrt{2m_c\Omega}$, that is, photons and excitons remain coupled as long as the kinetic energy of the photons is of the order of magnitude of the Rabi coupling $k^2/2m_c\sim \Omega$. Comparing against the typical momentum involved in the formation of the bound state $k_{scatt}\sim \sqrt{2m_X {\epsilon_B}} $ we obtain $k_{\Omega}/k_{scatt}\sim 10^{-3},$ and thus, the light-matter coupling does not change the scattering matrix for the spanned parameter space. 

We stress that although the light-matter coupling does not alter the scattering matrix, the dramatic changes in the polariton energies and their residues lead to substantial differences in the USC trion-polaritons compared to their SC counterpart, as the relevant energies at which the scattering matrix is evaluated does depend on the polariton energies.

\begin{figure*}[ht]
\includegraphics[width=.4\columnwidth]{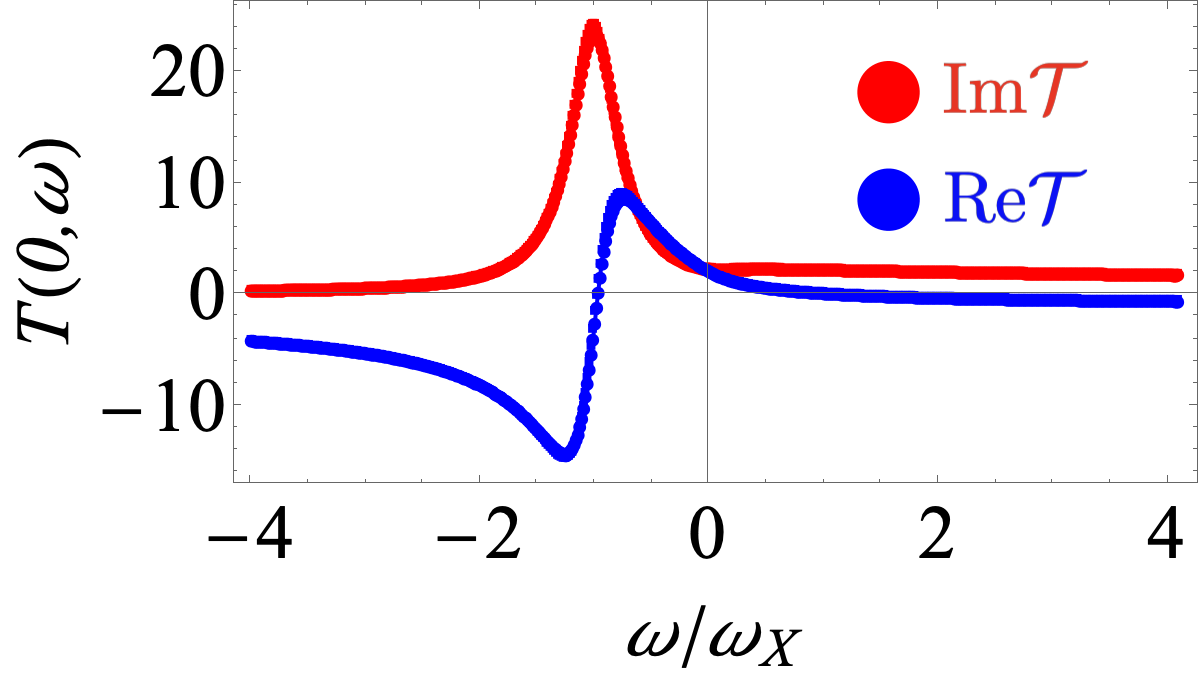}
\caption{Real (blue dots) and negative imaginary (red dots) parts of the scattering matrix as a function of frequency $\omega$ for $\Delta=0$ and $\Omega/\omega_X=0.5$, within the USC regime. The solid blue and red lines illustrate the scattering matrix without light-matter coupling.}
    \label{fig:s3}
\end{figure*}

\end{widetext}

\end{document}